\documentclass[aps,pre,reprint]{revtex4-2}

\usepackage{lmodern}

\usepackage[T1]{fontenc}

\usepackage[utf8]{inputenc}

\usepackage{amsmath,amssymb}

\everymath{\displaystyle}

\usepackage{siunitx}

\usepackage{physics}

\usepackage{color}

\makeatletter
\def\alain#1{\textcolor{black}{#1}} 
\makeatother

\makeatletter\let\NAT@sort\z@\makeatother

\begin{document}

\title{A formalism for the long-distance magnetic field generated by populations of neurons}

\author{Alexis O. \surname{García Rodríguez}}

\author{Alain Destexhe}

\affiliation{Paris-Saclay University, Institute of Neuroscience, CNRS, Gif sur Yvette, France}

\begin{abstract}

Brain activity can be measured using magnetic fields \alain{located at some distance from the brain}, a technique called magneto-encephalography (MEG). The origin of such magnetic fields are the ionic currents involved in neuronal activity. While these processes are well known at the microscopic scale, it is less clear how large populations of neurons generate magnetic fields. Here, we attempt to make such a link by deriving a mean-field theory of magnetic field generation by populations of neurons. \alain{We use the concept of current dipoles to build a theory that can be applied to populations of neurons.  We also evaluate the contributions of neuronal current sources such as the synaptic currents or action potential currents.  This theory should be useful to calculate the magnetic field located at long distances from neuronal populations, such as in MEG.}

\end{abstract}

\maketitle

\section*{Introduction}

\alain{Magnetic field measurements constitute a very attractive method to measure neuronal activity, in particular through magneto-encephalography (MEG), which measures the magnetic field at some distance from the brain surface, as reviewed previously~\cite{hamalainen,gulrajani}.  Contrary to the genesis of electric fields and potentials,} membrane currents do not contribute much to extracellular magnetic fields, because they cancel due to the cylindrical shape of neuronal cables, but that the main contribution to the extracellular magnetic field is the axial current \alain{in neuronal dendrites}~\cite{hamalainen}. It was shown previously that the ``microscopic'' magnetic fields measured for muscle fibers using miniaturized sensors, could be quantitatively matched by computational models taking into account the axial currents~\cite{barbieri}.

\alain{Another difference between electric and magnetic fields is that the electric potential around neurons crucially depends on the electric properties of the extracellular medium, while magnetic fields are believed to have negligible interactions with the extracellular medium~\cite{gulrajani}.  Note that some studies have found that the electric properties of extracellular media also affect magnetic fields, because they influence the neuronal current sources that produce the field~\cite{bedard}.  We will assume here that there is no such dependence and will consider neurons embedded in resistive (Ohmic) media, with no capacitive or diffusive effects.}

\alain{There exists well established theories to calculate the microscopic currents generating extracellular electric potentials, such as the neuronal cable theory~\cite{Rall95}.  Such theories can be formulated macroscopically, using a mean-field formulation of Maxwell equations~\cite{bedard2013}, yielding expressions for the extracellular potential at scales larger than the variations of resistivity and permittivity.  Such theories can easily integrate the influence of multiple neuronal current sources, and thus can be used to calculate the electric potential resulting from populations of neurons.}

However, how to account for ``macroscopic'' neuronal magnetic fields is less clear. One of the reasons is that the vectorial nature of the magnetic field will lead to partial cancellation of the fields generated by single neurons, in a way that critically depends on their orientation. Taking advantage of the fact that the principal neurons in cerebral cortex form electric dipoles, which are arranged in parallel, we derive here a mean-field formulation to account for macroscopic neuronal magnetic fields.  \alain{We start from the notion of current dipole, which we contrast with the notion of electric dipole, showing that they constitute distinct concepts with their own definition of the dipole moment.  We next evaluate the contribution to different types of current sources, such as the synaptic currents and action potential currents, to generate the magnetic field located far away (centimenters) from neurons, such as typically in MEG measurements.} 

\section{Current dipole}

A current line segment can be seen as a current dipole. This is a combination of a current sink and a current source of equal magnitude~\cite{hobbie}. The current enters at one end---a current sink---and leaves the segment at the other end---a current source. Synaptic and action potential currents can be described as a set of tiny current dipoles.

In the \alain{case of synaptic currents,} we have a single current dipole \alain{at} each synapse. Here the current either flows into \emph{or} out of the cell. Transmitter molecules in the synapse change the membrane permeability for specific ions. In the excitatory (depolarization) case, \(\mathrm{Na}^+\) channels open and current flows into the cell. For inhibitory (hyperpolarization) synaptic currents, \(\mathrm K^+\) or \(\mathrm{Cl}^-\) channels open and current flows out of the cell.

Action potential currents, in turn, can be seen as two oppositely oriented current dipoles forming a current quadrupole~\cite{hamalainen}. Here two currents flow in opposite directions. \(\mathrm{Na}^+\) ions go into the cell causing the membrane to depolarize and \(\mathrm K^+\) ions move out of the cell causing repolarization. These currents produce a change in the membrane potential that lasts about \SI{1}{ms}, a time scale \(10\) times smaller than that of a postsynaptic potential lasting tens of \si{ms}~\cite{hamalainen}.

The opposite currents creating an action potential produce magnetic fields that tend to cancel each other. Hence a relatively small magnetic field, compared to that of synaptic currents, is generated in that case. The brain magnetic field~\footnote{More precisely, we mean here the magnitude of the magnetic induction \(\vb*B\), called also magnetic flux density, which is measured in MEG recordings. In a linear and isotropic medium with magnetic permeability \(\mu\) we have \(\vb*H=\vb*B/\mu\), where \(\vb*H\) is the magnetic field} is typically up to \SI{5e-13}{T}~\cite{hamalainen} which is still very small compared to the earth magnetic field of about \SI{5e-5}{T}~\cite{griffithsbook}. Superconducting quantum interference device (SQUID) detectors are employed to record the MEG signals. These sensors work at liquid helium temperature (about \SI{4}{K}=\SI{-269}{\celsius}) and are placed some centimeters away from the scalp.

Our goal is to obtain the long-distance magnetic field produced by ionic currents in small regions containing populations of neurons. The synaptic currents in each region define an overall (time-dependent) current dipole. The dimensions of each region must be small compared to the average distance from the region to the measurement point. Considering this distance in the order of centimeters, the region size can be thought in the order of millimeters. One of those relatively small regions can then have a volume of about \SI{1}{mm^3}. We can take the values of the cortical surface, the average cortical thickness, and the total number of neurons in the human cerebral cortex, given by about \SI{1.8e5}{mm^2}, \SI{2.6}{mm}, and \num{16} billion neurons, respectively~\cite{essen}. We then find about \num{4.7e5} regions of \SI{1}{mm^3} in the human cerebral cortex, each one with an average estimate of about \num{3.4e4} neurons.

\subsection{Long-distance magnetic field of a current line segment}

The Biot--Savart law for a time-independent line current
\begin{equation}
\vb*B(\vb*r)=\frac\mu{4\pi}\,\mathcal I\int\dd{\vb*l'}\cp\frac{\vb*r-\vb*r'}{\vqty{\vb*r-\vb*r'}^3},
\label{biotsavartBI}
\end{equation}
can be employed to estimate the magnetic induction \(\vb*B_{seg}(\vb*r)\) of a current line segment at a relatively long distance. This leads to the concept of current dipole moment. The measurement point is located at \(\vb*r\). We consider a linear and isotropic medium with the same magnetic permeability \(\mu\) throughout space. \(\mathcal I\) is the current and \(\dd{\vb*l'}\) is a vector element of length in the current flow direction. \(\vb*r'\) refers to a given point in the segment.

Given a long distance to the point of interest when compared to the size of the current line segment, any point \(\vb*r_0^{\mathstrut}\) in it can be used to characterize its average position. We have that \(\vqty{\vb*r'-\vb*r_0^{\mathstrut}}\ll\vqty{\vb*r-\vb*r_0^{\mathstrut}}\) for every point in the segment. The long-distance magnetic field can be estimated by considering in Eq.~\eqref{biotsavartBI} all \(\vb*r'\approx\vb*r_0^{\mathstrut}\). We then have for the given segment
\begin{equation}
\vb*B_{seg}(\vb*r)\approx\frac\mu{4\pi}\frac{\vb*{\mathcal Q}_{seg}\cp\vu*R\pqty{\vb*r, \vb*r_0^{\mathstrut}}}{\vqty{\vb*R\pqty{\vb*r, \vb*r_0^{\mathstrut}}}^2},
\label{Bseg}
\end{equation}
where
\begin{equation}
\vu*R\pqty{\vb*r, \vb*r_0^{\mathstrut}}=\frac{\vb*R\pqty{\vb*r, \vb*r_0^{\mathstrut}}}{\vqty{\vb*R\pqty{\vb*r, \vb*r_0^{\mathstrut}}}}\qc\vb*R\pqty{\vb*r, \vb*r_0^{\mathstrut}}=\vb*r-\vb*r_0^{\mathstrut},
\end{equation}
and
\begin{equation}
\vb*{\mathcal Q}_{seg}=\mathcal I\int\limits_{\mathcal L_{seg}}\dd{\vb*l'}.
\label{momentseg}
\end{equation}
Here \(\vb*{\mathcal Q}_{seg}\) is the current dipole moment of the line segment \(\mathcal L_{seg}\). Its magnitude is given by the product of the current times the distance between the ends of the segment. This is similar to an electric dipole moment which is given by the product of the positive charge times the distance between the charges forming the dipole.

Eq.~\eqref{Bseg} shows a decrease with the square of the distance for the magnetic field of a current dipole. This behavior can be expected for synaptic currents due to their dipolar nature~\cite{hamalainen}. Below we reproduce that behavior and relate the synaptic current dipole moment of a region to the currents driven by the electric field.

Later we study the long-distance magnetic field for action potential currents. It decreases faster---with the cube of the distance---than in the synaptic case. This is characteristic for the quadrupolar case~\cite{hamalainen}. We derive such a decrease by approximating to zero the current dipole moment in this case and then use the \emph{magnetic} dipole moment for the action potential currents.

\subsection{Current dipole moment for volume currents}

The current dipole moment \(\vb*{\mathcal Q}_{reg}\) of a given volume region \(\mathcal V_{reg}\) can be expressed in terms of the current density \(\vb*{\mathcal J}(\vb*r')\) at the different points \(\vb*r'\) in the region. Considering \eqref{momentseg} we write
\begin{equation}
\vb*{\mathcal Q}_{reg}=\int\limits_{\mathcal V_{reg}}\delta\vb*{\mathcal Q}\qc\delta\vb*{\mathcal Q}=\mathcal I'\dd{\vb*l'}=\mathcal J(\vb*r')\dd{s'}\dd{\vb*l'},
\end{equation}
where \(\dd{s'}\) is a surface element perpendicular to the current flow direction at point \(\vb*r'\). Then
\begin{equation}
\delta\vb*{\mathcal Q}=\mathcal J(\vb*r')\dd{s'}\dd{\vb*l'}=\vb*{\mathcal J}(\vb*r')\dd{s'}\dd{l'}=\vb*{\mathcal J}(\vb*r')\dd[3]{r'},
\end{equation}
where \(\dd[3]{r'}\) is a volume element. The current density \(\vb*{\mathcal J}(\vb*r')\) is therefore the density per unit volume, at point \(\vb*r'\), of the current dipole moment. We thus have
\begin{equation}
\vb*{\mathcal Q}_{reg}=\int\limits_{\mathcal V_{reg}}\vb*{\mathcal J}(\vb*r')\dd[3]{r'}.
\label{gendipm}
\end{equation}

\section{Biot--Savart law and the nonstatic case}

\subsection{\alain{The classic Biot-Savart law and its generalization}}

The Biot--Savart law for volume currents is
\begin{equation}
\vb*B(\vb*r)=\frac\mu{4\pi}\int\vb*J(\vb*r')\cp\frac{\vb*r-\vb*r'}{\vqty{\vb*r-\vb*r'}^3}\dd[3]{r'},
\label{biotsavartBJ}
\end{equation}
where the integration is over all space and \(\vb*J\) is the free current density which is assumed to be confined to a finite region. As mentioned earlier, we consider a linear and isotropic medium with uniform permeability \(\mu\). Eq.~\eqref{biotsavartBJ} can be proved to be valid in the static case where both \(\vb*J\) and the free charge density \(\rho\) are time-independent. It can be shown~\cite{griffithsbook} that \(\vb*B(\vb*r)\) given by \eqref{biotsavartBJ} satisfies the Ampère’s law in differential form
\begin{equation}
\curl\vb*B(\vb*r)=\mu\vb*J(\vb*r),
\label{Amp}
\end{equation}
provided
\begin{equation}
\divergence\vb*J(\vb*r)=0.
\label{divJ}
\end{equation}
This characterizes the magnetostatic case where \(\rho\) is independent of time \(t\), i.e., \(\pdv*\rho{t}=0\). The continuity equation \(\divergence\vb*J+\pdv*\rho{t}=0\), expressing charge conservation, becomes Eq.~\eqref{divJ}.

On the other hand, the brain tissue represents a nonstatic situation and the use of the Biot--Savart law must be justified. Here the charge density, given by the ion concentrations, depends on time at the different points in space. As noted below, Eq.~\eqref{biotsavartBJ} still holds in the special nonstatic case where the free charge density changes linearly with time~\cite{griffiths}. The continuity equation here implies that the free current density does not depend on time and the same happens with \(\vb*B\) as shown by Eq.~\eqref{biotsavartBJ}. However, MEG measurements show time-varying magnetic signals necessarily corresponding to time-varying currents in the brain. In this connection, a time-dependent version of Eq.~\eqref{biotsavartBJ}, given by
\begin{equation}
\vb*B(\vb*r,t)=\frac\mu{4\pi}\int\vb*J(\vb*r',t)\cp\frac{\vb*r-\vb*r'}{\vqty{\vb*r-\vb*r'}^3}\dd[3]{r'},
\label{biotsavartBJt}
\end{equation}
could be proposed. Nevertheless, as noted also below, Eq.~\eqref{biotsavartBJt} is warranted in the particular case where \(\vb*J(\vb*r',t)\) is linear in time~\cite{griffiths}. Here \(\vb*B\), according to \eqref{biotsavartBJt}, is also linear in time which neither corresponds to the situation in MEG where oscillating magnetic signals are recorded.

Therefore the general nonstatic case, where \(\rho\) and \(\vb*J\) depend arbitrarily on time, must be considered. Then, instead of \eqref{Amp}, the Ampère--Maxwell equation
\begin{equation}
\curl\vb*B(\vb*r,t)=\mu\bqty{\vb*J(\vb*r,t)+\pdv{\vb*D(\vb*r,t)}{t}},
\label{AmpMaxw}
\end{equation}
must be satisfied. Here \(\vb*D\) is the vector of electric displacement, such that
\begin{equation}
\divergence\vb*D(\vb*r,t)=\rho(\vb*r,t).
\label{divD}
\end{equation}
The continuity equation
\begin{equation}
\divergence\vb*J(\vb*r,t)+\pdv{\rho(\vb*r,t)}{t}=0,
\label{conteq}
\end{equation}
being a general statement of charge conservation, remains valid in this case. It implies, using Eq.~\eqref{divD}, that
\begin{equation}
\divergence[\vb*J(\vb*r,t)+\pdv{\vb*D(\vb*r,t)}{t}]=0,
\label{divJMaxwterm}
\end{equation}
instead of Eq.~\eqref{divJ}. This can also be obtained directly from Eq.~\eqref{AmpMaxw}. The terms inside the brackets form the generalized current density used previously to include charge accumulation in cable equations~\cite{bedard,bedard2013}.

Aiming to obtain a time-dependent generalization of the Biot--Savart law, the Maxwell's equation \(\divergence\vb*B(\vb*r,t)=0\) can be considered. This allows us to introduce the vector potential \(\vb*A(\vb*r,t)\), so that
\begin{equation}
\vb*B(\vb*r,t)=\curl\vb*A(\vb*r,t).
\label{Bvectpot}
\end{equation}
Substituting this into Eq.~\eqref{AmpMaxw} results in a Biot--Savart-like expression~\cite{[][{. The derivation given in this reference for free space similarly applies when considering a linear and isotropic medium with uniform permeability.}]weber,jefimenko}, which reads
\begin{equation}
\begin{aligned}
\vb*B(\vb*r,t)=\frac\mu{4\pi}\int&\bqty{\vb*J(\vb*r',t)+\pdv{\vb*D(\vb*r',t)}{t}}\\
&\cp\frac{\vb*r-\vb*r'}{\vqty{\vb*r-\vb*r'}^3}\dd[3]{r'}.
\end{aligned}
\label{BJD}
\end{equation}
This satisfies Eq.~\eqref{AmpMaxw} by taking into account Eq.~\eqref{divJMaxwterm} and reduces to the Biot--Savart law \eqref{biotsavartBJ} in the static case.

From a practical point of view, however, Eq.~\eqref{BJD} is not helpful as it turns out to be self-referential~\cite{jefimenko,griffiths}. Eq.~\eqref{BJD} requires the electric displacement \(\vb*D\) to be known at all points of space. Considering a linear and isotropic medium with permittivity \(\epsilon\), we have \(\vb*D=\epsilon\vb*E\), where \(\vb*E\) is the vector of electric field. The latter, meanwhile, is determined by \(\vb*B\) through the Faraday's law of induction, \(\curl\vb*E(\vb*r,t)=-\pdv*{\vb*B(\vb*r,t)}{t}\). Therefore, in order to evaluate \(\vb*B\) by employing Eq.~\eqref{BJD}, we must know \(\vb*B\) itself at all points of space.

Eq.~\eqref{BJD} cannot thus be actually used to obtain \(\vb*B\) for time-dependent systems. It requires to know the displacement current density \(\pdv*{\vb*D}{t}\) which has \(\vb*B\) as its own source~\cite{jefimenko}. We employ below the proper general solution given to this problem.

\subsection{Jefimenko's equation: The general nonstatic case}

A suitable time-dependent generalization of the Biot--Savart law was derived by Jefimenko~\cite{[][{; 2nd ed. (Electret Scientific, Star City, 1989).}]jefimenkobook,jefimenko,griffithsbook}, given by
\begin{equation}
\begin{aligned}
\hspace{-1.7pt}\vb*B(\vb*r,t)=\frac\mu{4\pi}\int&\bqty{\vb*J(\vb*r',t_{ret})+\frac{\vqty{\vb*r-\vb*r'}}v\pdv{\vb*J(\vb*r',t_{ret})}{t}}\\
&\cp\frac{\vb*r-\vb*r'}{\vqty{\vb*r-\vb*r'}^3}\dd[3]{r'}.
\end{aligned}
\label{Jefmko}
\end{equation}
An infinite, linear and isotropic medium with uniform permittivity \(\epsilon\) and permeability \(\mu\) is assumed. Here
\begin{equation}
t_{ret}=t-\vqty{\vb*r-\vb*r'}/v,
\label{tret}
\end{equation}
is the retarded time, and
\begin{equation}
v=1/\sqrt{\epsilon\mu},
\label{veloc}
\end{equation}
is the velocity of light in the medium. Eq.~\eqref{Jefmko} can be obtained from \eqref{Bvectpot} with
\begin{equation}
\vb*A(\vb*r,t)=\frac\mu{4\pi}\int\frac{\vb*J(\vb*r',t_{ret})}{\vqty{\vb*r-\vb*r'}}\dd[3]{r'}.
\end{equation}
The current sources in the vector potential must thus be evaluated at the retarded time \eqref{tret}. The status of the sources at this time determines \(\vb*B\) at the later time \(t\). It takes a time \(\vqty{\vb*r-\vb*r'}/v\) for an electromagnetic signal to travel from a source at point \(\vb*r'\) to the point \(\vb*r\) where \(\vb*B\) is measured. In this sense, Eq.~\eqref{Jefmko} is a causality-based general solution of Maxwell's equations.

Eq.~\eqref{Jefmko} simplifies to the static form \eqref{biotsavartBJ} of the Biot--Savart law in the case of a time-independent current density. This does not require the free charge density to be time-independent as well but, according to the continuity equation \eqref{conteq}, it may change linearly with time~\cite{griffiths}. It can also be verified that Eq.~\eqref{Jefmko} takes the non-retarded form \eqref{biotsavartBJt} when the current density is linear in time~\cite{griffiths}, that is, when \(\vb*J(\vb*r',t)=\vb*J(\vb*r',0)+t\mkern1mu\vb*C(\vb*r')\), where \(\vb*C(\vb*r')\) is a time-independent function.

\subsection{Magneto-quasistatic approximation}

It is worth considering the Jefimenko's equation \eqref{Jefmko} in the quasistatic case where the current density \(\vb*J(\vb*r',t)\), as well as the corresponding charge density, changes relatively slowly with time~\cite{griffiths}. The characteristic time for this change to occur must be large compared to the signal traveling time \(t_{trav}(\vb*r,\vb*r')=\vqty{\vb*r-\vb*r'}/v\). Considering \eqref{veloc} we have
\begin{equation}
t_{trav}(\vb*r,\vb*r')=\vqty{\vb*r-\vb*r'}\sqrt{\epsilon\mu}.
\label{ttrav}
\end{equation}
Similarly to the static case, the current density \(\vb*J\) in Eq.~\eqref{Jefmko} is assumed to be confined to a finite region of space. The source point \(\bar{\vb*r}'\) which is farthest from the field point \(\vb*r\) determines the largest traveling time \(t_{trav}(\vb*r,\bar{\vb*r}')\).

In the quasistatic case it is useful to expand the current density \(\vb*J(\vb*r',t_{ret})=\vb*J\bqty{\vb*r',t-t_{trav}(\vb*r,\vb*r')}\) in Eq.~\eqref{Jefmko} as a Taylor's series, so that
\begin{equation}
\begin{aligned}
\vb*J(\vb*r',t_{ret})&=\vb*J(\vb*r',t)-t_{trav}(\vb*r,\vb*r')\pdv{\vb*J(\vb*r',t)}{t}\\
&\quad+\frac12t_{trav}^2(\vb*r,\vb*r')\pdv[2]{\vb*J(\vb*r',t)}{t}+\cdots.
\end{aligned}
\end{equation}
This leads to
\begin{equation}
\begin{aligned}
\hspace{-.4pt}\vb*B(\vb*r,t)=\frac\mu{4\pi}\int&\left[\vb*J(\vb*r',t)-\frac12t_{trav}^2(\vb*r,\vb*r')\pdv[2]{\vb*J(\vb*r',t)}{t}\right.\\
&\left.\vphantom{\pdv[2]{\vb*J(\vb*r',t)}{t}}+\cdots\right]\cp\frac{\vb*r-\vb*r'}{\vqty{\vb*r-\vb*r'}^3}\dd[3]{r'}.
\end{aligned}
\label{JefmkoExp}
\end{equation}

The time dependency of the current density, and correspondingly that of \(\vb*B\), can be described through the different frequency components. A component of angular frequency \(w\) corresponds to a time oscillation with period \(T(w)=2\pi/w\). Here it is helpful to notice that the main contributions in neuromagnetism are from frequencies below a typical value \(f_{max}=\SI{100}{Hz}\)~\cite{hamalainen}. We thus consider a maximum angular frequency
\begin{equation}
w_{max}=2\pi f_{max},
\label{wmax}
\end{equation}
and write
\begin{equation}
\vb*J(\vb*r',t)\approx\int\limits_0^{w_{max}} e^{iwt}\skew{6}\widetilde{\vb*J}(\vb*r',w)\dd w.
\label{Jws}
\end{equation}
Substituting this into Eq.~\eqref{JefmkoExp} yields
\begin{equation}
\vb*B(\vb*r,t)\approx\int\limits_0^{w_{max}} e^{iwt}\skew{4}\widetilde{\vb*B}(\vb*r,w)\dd w,
\label{Bws}
\end{equation}
where
\begin{equation}
\begin{aligned}
\skew{4}\widetilde{\vb*B}(\vb*r,w)=\frac\mu{4\pi}\int&\left[\skew{6}\widetilde{\vb*J}(\vb*r',w)+\frac12w^2t_{trav}^2(\vb*r,\vb*r')\right.\\
&\left.\vphantom{\frac12}\times\skew{6}\widetilde{\vb*J}(\vb*r',w)+\cdots\right]\\
&\cp\frac{\vb*r-\vb*r'}{\vqty{\vb*r-\vb*r'}^3}\dd[3]{r'}.
\end{aligned}
\label{Brw}
\end{equation}
In the quasistatic case we must have at every source point \(\vb*r'\)
\begin{equation}
w^2t_{trav}^2(\vb*r,\vb*r')\ll1,\qq{for all}0<w<w_{max}.
\end{equation}
Since \(wt_{trav}(\vb*r,\vb*r')\leqslant w_{max}t_{trav}(\vb*r,\bar{\vb*r}')\), it is enough to require that
\begin{equation}
w_{max}^2t_{trav}^2(\vb*r,\bar{\vb*r}')\ll1.
\label{condgensol}
\end{equation}
This describes the quasistatic case through the current frequency component with the shortest oscillation period \(2\pi/w_{max}=1/f_{max}\). This time must be large enough compared to the largest traveling time \(t_{trav}(\vb*r,\bar{\vb*r}')\). In that case, Eq.~\eqref{Brw} can be replaced by
\begin{equation}
\skew{4}\widetilde{\vb*B}(\vb*r,w)\approx\frac\mu{4\pi}\int\skew{6}\widetilde{\vb*J}(\vb*r',w)\cp\frac{\vb*r-\vb*r'}{\vqty{\vb*r-\vb*r'}^3}\dd[3]{r'},
\end{equation}
for all \(0<w<w_{max}\). Using \eqref{Bws} and \eqref{Jws}, we then have
\begin{equation}
\vb*B(\vb*r,t)\approx\frac\mu{4\pi}\int\vb*J(\vb*r',t)\cp\frac{\vb*r-\vb*r'}{\vqty{\vb*r-\vb*r'}^3}\dd[3]{r'}.
\label{biotsavartBJtapprox}
\end{equation}
This is similar to the static form \eqref{biotsavartBJ} of the Biot--Savart law. Expression \eqref{biotsavartBJtapprox} is the quasistatic approximation of the general solution \eqref{Jefmko} of Maxwell's equations. It is valid provided \eqref{condgensol} is satisfied.

We note that a similar expression to \eqref{condgensol} is well known for the magneto-quasistatic approximation of Maxwell's equations~\cite{[][{; see also supplementary material. The discussion here is based on J. R. Melcher, \emph{Continuum Electromechanics} (MIT Press, Cambridge, 1981).}]wagner,haus}, as well as their electro-quasistatic approximation~\cite{wagner,haus,gratiy,*gratiyadd}. In these cases, however, instead of the traveling distance \(\vqty{\vb*r-\bar{\vb*r}'}\) in Eq.~\eqref{ttrav}, from the most distant source point \(\bar{\vb*r}'\) to the field point \(\vb*r\), one typical or characteristic length scale of the medium is employed. The system dimensions are assumed to be all within a factor of two or so of each other~\cite{haus}. The typical length is then used to normalize~\cite{wagner} or approximate~\cite{haus} all spatial derivatives in Maxwell's equations. In gray matter, the typical length is approximated by the cortical thickness~\cite{wagner,gratiy}. The approaches in Refs.~\citenum{wagner,haus,gratiy} are not based on the general solution of Maxwell's equations. This is given in the magnetic case by Eq.~\eqref{Jefmko}. As shown above, this leads to expression \eqref{condgensol} and therefore, to the traveling distance \(\vqty{\vb*r-\bar{\vb*r}'}\) as the appropriate length to define the quasistatic dynamics.

\subsection{The quasistatic case of the brain}

In the following we check whether the requirement \eqref{condgensol} is fulfilled in the specific case of the brain. According to Eqs.~\eqref{wmax} and \eqref{ttrav}, we must have
\begin{equation}
(2\pi f_{max})^2\vqty{\vb*r-\bar{\vb*r}'}^2\epsilon\mu\ll1.
\end{equation}
As mentioned above, we typically have \(f_{max}=\SI{100}{Hz}\)~\cite{hamalainen}. We consider the distance \(\vqty{\vb*r-\bar{\vb*r}'}\) mainly given by the average size of the human brain of about \SI{15}{cm}. This is much larger than the cortical thickness of the order of \SI{1}{mm}. The distance of about few centimeters between sensors and the scalp is also relatively small. We then take \(\vqty{\vb*r-\bar{\vb*r}'}\approx\SI{.1}{m}\). With regard to the permittivity, we have \(\epsilon\approx\num{e7}\epsilon_0^{\mathstrut}\) for gray matter at frequencies below \SI{100}{Hz}~\cite{wagner,gabriel}. Here \(\epsilon_0^{\mathstrut}\approx\SI{8.85e-12}{C^2/(N.m^2)}\) is the permittivity of free space. A magnetic permeability close to that of free space is typical for biological tissues~\cite{wagner}. We then assume \(\mu\approx\mu_0^{\mathstrut}\approx\SI{1.26e-6}{N/A^2}\), and obtain~\footnote{It can be seen that using \(f_{max}=\SI{10}{kHz}\) and a characteristic length of \SI{1}{mm}, corresponding to the order of the cortical thickness, yields also the value \(\num{4e-7}\) for the electro-quasistatic approximation in Ref.~\citenum{gratiy}}
\begin{equation}
(2\pi f_{max})^2\vqty{\vb*r-\bar{\vb*r}'}^2\epsilon\mu\approx\num{4e-7}\ll1.
\end{equation}
The condition \eqref{condgensol} is thus very well satisfied, indicating a quasistatic dynamics. Therefore, we take \eqref{biotsavartBJtapprox} as a valid expression to calculate \(\vb*B(\vb*r,t)\).

In closing this section, we discuss about a condition which is different from \eqref{condgensol} and that has been proposed in Ref.~\citenum{hamalainen} to justify the validity of expression \eqref{biotsavartBJtapprox} in the quasistatic case. The analysis in Ref.~\citenum{hamalainen} is based on the Ampère--Maxwell equation, given here by Eq.~\eqref{AmpMaxw}. The frequency components of the displacement term \(\pdv*{\vb*D(\vb*r,t)}{t}\) are compared to those of the ohmic part of the current density \(\vb*J(\vb*r,t)\). They are given, respectively, by \(iw\epsilon\skew{4}\widetilde{\vb*E}(\vb*r,w)\) and \(\sigma\skew{4}\widetilde{\vb*E}(\vb*r,w)\), where \(\skew{4}\widetilde{\vb*E}(\vb*r,w)\) is the frequency component of the electric field, and \(\sigma\) is the electrical conductivity of the medium. The ratio between the absolute values of those frequency components leads then to the condition
\begin{equation}
w_{max}\tau_{MW}^{\mathstrut}\ll1,
\label{condAmpMaxw}
\end{equation}
where \(\tau_{MW}^{\mathstrut}=\epsilon/\sigma\) is the Maxwell--Wagner reaction time.

According to \eqref{condAmpMaxw}, the shortest oscillation period \(2\pi/w_{max}=1/f_{max}\) must be large enough compared to the Maxwell--Wagner time. If that is the case, then the displacement term can be neglected and the Ampère--Maxwell equation can be replaced by \(\curl\vb*B(\vb*r,t)\approx\mu\vb*J(\vb*r,t)\). This is similar to the Ampère’s law \eqref{Amp} describing the static case and results in the expression \eqref{biotsavartBJtapprox}, as can be seen from Eq.~\eqref{BJD}.

The expressions \eqref{condAmpMaxw} and \eqref{condgensol} are clearly different from one another. In the first place, one is linear while the other is quadratic and secondly, they define the quasistatic dynamics in terms of essentially different times. These are the Maxwell--Wagner reaction time and the traveling time which also differ quantitatively. Using \(\epsilon\approx\num{e7}\epsilon_0^{\mathstrut}\) as before~\cite{wagner,gabriel}, and \(\sigma\approx\SI{.3}{\ohm^{-1}.m^{-1}}\) for gray matter at frequencies below \SI{100}{Hz}~\cite{wagner,hamalainen}, we have \(\tau_{MW}^{\mathstrut}\approx\SI{3e-4}{s}\). Taking \(\vqty{\vb*r-\bar{\vb*r}'}\approx\SI{.1}{m}\), \(\epsilon\approx\num{e7}\epsilon_0^{\mathstrut}\), and \(\mu\approx\mu_0^{\mathstrut}\), we have \(t_{trav}(\vb*r,\bar{\vb*r}')=\vqty{\vb*r-\bar{\vb*r}'}\sqrt{\epsilon\mu}\approx\SI{1}{\us}\). The expression \eqref{condAmpMaxw} thus requires a relatively slower dynamics than in the case of expression \eqref{condgensol}.

Now we evaluate the fulfillment of the requirement \eqref{condAmpMaxw}. We use the typical value \(f_{max}=\SI{100}{Hz}\)~\cite{hamalainen} and \(\tau_{MW}^{\mathstrut}\approx\SI{3e-4}{s}\) which was obtained before. This yields
\begin{equation}
w_{max}\tau_{MW}^{\mathstrut}=2\pi f_{max}\tau_{MW}^{\mathstrut}\approx\num{.2}.
\end{equation}
The condition \eqref{condAmpMaxw} is thus minimally satisfied. We note that a permittivity \(\epsilon\approx\num{e5}\epsilon_0^{\mathstrut}\) was considered in Ref.~\citenum{hamalainen} giving \(w_{max}\tau_{MW}^{\mathstrut}\approx\num{2e-3}\ll1\). However, according to Refs.~\citenum{wagner} and \citenum{gabriel}, \(\num{e5}\epsilon_0^{\mathstrut}\) is an underestimated value of the permittivity for gray matter at frequencies below \SI{100}{Hz}.

The fact that the condition \eqref{condAmpMaxw} is minimally fulfilled indicates that it is not a very good approximation to neglect the displacement term in the Ampère--Maxwell equation. That is the case according to typical values of frequency, permittivity and conductivity in the brain. Neglecting the displacement term is not a safe option and the general solution of Maxwell's equations, given above by Eq.~\eqref{Jefmko}, must be considered.

The expression \eqref{condAmpMaxw}, which corresponds to neglecting the displacement term in the Ampère--Maxwell equation, is not a necessary condition to justify the validity of expression \eqref{biotsavartBJtapprox} in the quasistatic case. As pointed out after Eq.~\eqref{BJD}, the electric displacement \(\vb*D=\epsilon\vb*E\) depends on \(\vb*B\) itself through the Faraday's law of induction, \(\curl\vb*E(\vb*r,t)=-\pdv*{\vb*B(\vb*r,t)}{t}\). Therefore the displacement current density \(\pdv*{\vb*D}{t}\) must not be considered as a source of \(\vb*B\) in the conventional sense~\cite{jefimenko}. The time-dependent generalization of the Biot--Savart law, Eq.~\eqref{Jefmko}, does not contain indeed the displacement current density but only \(\vb*J\). Eq.~\eqref{Jefmko} introduces the signal traveling time \(\vqty{\vb*r-\vb*r'}/v\) in terms of which the quasistatic dynamics can be defined. Correspondingly, the Maxwell--Wagner time is suitable for a quasistatic approximation of the Ampère--Maxwell equation but not of the generalized Biot--Savart law \eqref{Jefmko}.

\section{Magnetic field of the different currents}

According to the previous section, we take \eqref{biotsavartBJtapprox} as a valid expression to calculate \(\vb*B(\vb*r,t)\). Considering the typical value \(\mu\approx\mu_0^{\mathstrut}\)~\cite{wagner}, it becomes
\begin{equation}
\vb*B(\vb*r,t)\approx\frac{\mu_0^{\mathstrut}}{4\pi}\int\vb*J(\vb*r',t)\cp\frac{\vb*r-\vb*r'}{\vqty{\vb*r-\vb*r'}^3}\dd[3]{r'}.
\label{biotsavartBJtapproxmu0}
\end{equation}

The current density \(\vb*J\) can be written as the sum of two components~\cite{geselowitz67}. First, the conduction current density
\begin{equation}
\vb*J^c(\vb*r,t)=\sigma(\vb*r,t)\vb*E(\vb*r,t),
\end{equation}
which corresponds to the current caused by the electric field \(\vb*E\). This is produced by the charge carriers in the conducting medium. In the electro-quasistatic approximation~\cite{gratiy} we have \(\curl\vb*E(\vb*r,t)=-\pdv*{\vb*B(\vb*r,t)}{t}\approx\vb0\). This allows to use the scalar potential \(\Phi(\vb*r,t)\) so that \(\vb*E(\vb*r,t)\approx-\grad\Phi(\vb*r,t)\). Therefore we have
\begin{equation}
\vb*J^c(\vb*r,t)\approx-\sigma(\vb*r,t)\grad\Phi(\vb*r,t).
\label{condcurrentdens}
\end{equation}
Here the medium is described as a nonhomogeneous conductor with conductivity \(\sigma(\vb*r,t)\) which is modeled on cellular scale. On the same microscopic, micrometer scale, we have a second component of the current density corresponding to the impressed currents~\cite{ilmoniemi} and represented by \(\vb*J^i(\vb*r,t)\). This stands for the current density which is not produced by the electric field \(\vb*E\) but mainly, by gradients of concentration in diffusion processes and electromotive chemical reactions in the cell pump mechanism. Then
\begin{equation}
\vb*J(\vb*r,t)=\vb*J^c(\vb*r,t)+\vb*J^i(\vb*r,t),
\label{currentdenscondimp}
\end{equation}
with the conduction current density \(\vb*J^c(\vb*r,t)\) given by \eqref{condcurrentdens}.

We assume the impressed current density mainly given by a current density \(\vb*J^s(\vb*r,t)\) corresponding to synaptic currents, and a current density \(\vb*J^a(\vb*r,t)\) corresponding to action potential currents. That is, we have
\begin{equation}
\vb*J^i(\vb*r,t)\approx\vb*J^s(\vb*r,t)+\vb*J^a(\vb*r,t).
\label{impcurrentdens}
\end{equation}
Thus
\begin{equation}
\vb*J(\vb*r,t)\approx\vb*J^c(\vb*r,t)+\vb*J^s(\vb*r,t)+\vb*J^a(\vb*r,t),
\label{currentdenscondsynact}
\end{equation}
and the expression \eqref{biotsavartBJtapproxmu0} can be written in the form
\begin{equation}
\vb*B(\vb*r,t)\approx\vb*B^c(\vb*r,t)+\vb*B^s(\vb*r,t)+\vb*B^a(\vb*r,t),
\end{equation}
where
\begin{equation}
\vb*B^c(\vb*r,t)=\frac{\mu_0^{\mathstrut}}{4\pi}\int\vb*J^c(\vb*r',t)\cp\frac{\vb*r-\vb*r'}{\vqty{\vb*r-\vb*r'}^3}\dd[3]{r'},
\label{biotsavartBJtapproxmu0cond}
\end{equation}
\begin{equation}
\vb*B^s(\vb*r,t)=\frac{\mu_0^{\mathstrut}}{4\pi}\int\vb*J^s(\vb*r',t)\cp\frac{\vb*r-\vb*r'}{\vqty{\vb*r-\vb*r'}^3}\dd[3]{r'},
\label{biotsavartBJtapproxmu0syn}
\end{equation}
and
\begin{equation}
\vb*B^a(\vb*r,t)=\frac{\mu_0^{\mathstrut}}{4\pi}\int\vb*J^a(\vb*r',t)\cp\frac{\vb*r-\vb*r'}{\vqty{\vb*r-\vb*r'}^3}\dd[3]{r'},
\label{biotsavartBJtapproxmu0act}
\end{equation}
correspond, respectively, to the conduction, synaptic and action potential currents.

We note that the case of conduction currents, given by expressions \eqref{biotsavartBJtapproxmu0cond} and \eqref{condcurrentdens}, has been solved exactly~\cite{geselowitz70}. Next we look at the synaptic and action potential cases.

\subsection{Long-distance magnetic field of synaptic currents}

We estimate the synaptic contribution \eqref{biotsavartBJtapproxmu0syn} by considering several relatively small regions in the medium, such that
\begin{equation}
\vb*B^s(\vb*r,t)=\sum_{reg}\vb*B_{reg}^s(\vb*r,t),
\label{Btsynallregs}
\end{equation}
where
\begin{equation}
\vb*B_{reg}^s(\vb*r,t)=\frac{\mu_0^{\mathstrut}}{4\pi}\int\limits_{\mathcal V_{reg}}\vb*J^s(\vb*r',t)\cp\frac{\vb*r-\vb*r'}{\vqty{\vb*r-\vb*r'}^3}\dd[3]{r'},
\label{biotsavartBJtapproxmu0synreg}
\end{equation}
corresponds to a region of volume \(\mathcal V_{reg}\). The dimensions of each region must be small compared to the average distance from the region to the field point \(\vb*r\). Taking any point \(\vb*r_{reg}^{\mathstrut}\) in the region, we must have
\begin{equation}
\vqty{\vb*r'-\vb*r_{reg}^{\mathstrut}}\ll\vqty{\vb*r-\vb*r_{reg}^{\mathstrut}},
\end{equation}
for every region point \(\vb*r'\).

Eq.~\eqref{biotsavartBJtapproxmu0synreg} can be written in the form
\begin{equation}
\vb*B_{reg}^s(\vb*r,t)=\curl\vb*A_{reg}^s(\vb*r,t),
\label{Bvectpotregsyn}
\end{equation}
with the vector potential
\begin{equation}
\vb*A_{reg}^s(\vb*r,t)=\frac{\mu_0^{\mathstrut}}{4\pi}\int\limits_{\mathcal V_{reg}}\frac{\vb*J^s(\vb*r',t)}{\vqty{\vb*r-\vb*r'}}\dd[3]{r'}.
\end{equation}
We expand the factor \(1/\vqty{\vb*r-\vb*r'}\) around \(\vb*r'=\vb*r_{reg}^{\mathstrut}\). Up to the first order term, we have
\begin{equation}
\begin{aligned}
\frac1{\vqty{\vb*r-\vb*r'}}&\approx\frac1{\vqty{\vb*r-\vb*r_{reg}^{\mathstrut}}}\\
&\quad+\eval{\grad'\frac1{\vqty{\vb*r-\vb*r'}}}_{\vb*r'=\vb*r_{reg}^{\mathstrut}}\vdot\pqty{\vb*r'-\vb*r_{reg}^{\mathstrut}},
\end{aligned}
\label{invdistexp}
\end{equation}
and therefore
\begin{equation}
\begin{aligned}
\vb*A_{reg}^s(\vb*r,t)&\approx\frac{\mu_0^{\mathstrut}}{4\pi}\frac1{\vqty{\vb*r-\vb*r_{reg}^{\mathstrut}}}\int\limits_{\mathcal V_{reg}}\vb*J^s(\vb*r',t)\dd[3]{r'}\\
&\quad+\frac{\mu_0^{\mathstrut}}{4\pi\vqty{\vb*r-\vb*r_{reg}^{\mathstrut}}^3}\int\limits_{\mathcal V_{reg}}\vb*J^s(\vb*r',t)\\
&\qquad\times\bqty{\pqty{\vb*r-\vb*r_{reg}^{\mathstrut}}\vdot\pqty{\vb*r'-\vb*r_{reg}^{\mathstrut}}}\dd[3]{r'}.
\end{aligned}
\label{vectpotregexpsyn}
\end{equation}
Similarly to Eq.~\eqref{gendipm}, the quantity
\begin{equation}
\vb*Q_{reg}^s(t)=\int\limits_{\mathcal V_{reg}}\vb*J^s(\vb*r',t)\dd[3]{r'},
\label{momentsyn}
\end{equation}
in the first term of expression \eqref{vectpotregexpsyn}, defines the synaptic current dipole moment. Assuming this is nonzero vector, we approximate the vector potential by
\begin{equation}
\vb*A_{reg}^s(\vb*r,t)\approx\frac{\mu_0^{\mathstrut}}{4\pi}\frac{\vb*Q_{reg}^s(t)}{\vqty{\vb*r-\vb*r_{reg}^{\mathstrut}}}.
\end{equation}
Using then Eq.~\eqref{Bvectpotregsyn}, we have
\begin{equation}
\vb*B_{reg}^s(\vb*r,t)\approx\frac{\mu_0^{\mathstrut}}{4\pi}\frac{\vb*Q_{reg}^s(t)\cp\vu*R\pqty{\vb*r, \vb*r_{reg}^{\mathstrut}}}{\vqty{\vb*R\pqty{\vb*r, \vb*r_{reg}^{\mathstrut}}}^2},
\label{Bregsyn}
\end{equation}
where
\begin{equation}
\vu*R\pqty{\vb*r, \vb*r_{reg}^{\mathstrut}}=\frac{\vb*R\pqty{\vb*r, \vb*r_{reg}^{\mathstrut}}}{\vqty{\vb*R\pqty{\vb*r, \vb*r_{reg}^{\mathstrut}}}},
\label{relposunitvect}
\end{equation}
with
\begin{equation}
\vb*R\pqty{\vb*r, \vb*r_{reg}^{\mathstrut}}=\vb*r-\vb*r_{reg}^{\mathstrut}.
\label{relposvect}
\end{equation}
Expression \eqref{Bregsyn} shows a decrease with the square of the distance for synaptic currents which is characteristic for the dipolar case~\cite{hamalainen}.

\subsection{Synaptic current dipole moment}

We must find the synaptic current dipole moment, \(\vb*Q_{reg}^s(t)\), defined by \eqref{momentsyn}. Integrating the expression \eqref{currentdenscondsynact} of the current density yields
\begin{equation}
\vb*Q_{reg}(t)\approx\vb*Q_{reg}^c(t)+\vb*Q_{reg}^s(t)+\vb*Q_{reg}^a(t),
\label{currentdenscondsynactint}
\end{equation}
where
\begin{equation}
\vb*Q_{reg}(t)=\int\limits_{\mathcal V_{reg}}\vb*J(\vb*r',t)\dd[3]{r'},
\label{momenttot}
\end{equation}
is the total current dipole moment of the region. Correspondingly
\begin{equation}
\vb*Q_{reg}^c(t)=\int\limits_{\mathcal V_{reg}}\vb*J^c(\vb*r',t)\dd[3]{r'},
\label{momentcond}
\end{equation}
is the conduction current dipole moment, and
\begin{equation}
\vb*Q_{reg}^a(t)=\int\limits_{\mathcal V_{reg}}\vb*J^a(\vb*r',t)\dd[3]{r'},
\label{momentact}
\end{equation}
is the current dipole moment for action potential currents. We consider an approximately cylindrical shape for the neuronal cables. Then we have opposite transverse currents on opposite sides of the cables whose contributions, in the current dipole moment \eqref{momentact} for action potential currents, approximately cancel each other. We then take
\begin{equation}
\vb*Q_{reg}^a(t)\approx\vb0.
\label{momentactapprox}
\end{equation}
Thus, from expression \eqref{currentdenscondsynactint} we have
\begin{equation}
\vb*Q_{reg}^s(t)\approx\vb*Q_{reg}(t)-\vb*Q_{reg}^c(t).
\label{momentsynapprox}
\end{equation}

We assume quasistatic conditions such that, based on the continuity equation, \(\grad'\vdot\vb*J(\vb*r',t)=-\pdv*{\rho(\vb*r',t)}{t}\approx0\)~\cite{geselowitz67}. Then, provided the contribution of the impressed currents at the region boundary is small compared to that of the conduction currents~\footnote{Specifically, as can be seen from \protect\eqref{rprimpcurrentdenssurfintcompx}, this contribution of the impressed currents at the surface \(\mathcal S_{reg}\) bounding the region, is given by the components of \(\int\limits_{\mathcal S_{reg}}\vb*r'\bqty{\vb*J^i(\vb*r',t)\vdot\dd{\vb*S'}}\). This term should be added to expression \protect\eqref{momenttotapprox} and its equivalent \protect\eqref{momenttotapproxequiv} in a broader approach}, the total current dipole moment \eqref{momenttot} is given by
\begin{equation}
\vb*Q_{reg}(t)\approx\int\limits_{\mathcal S_{reg}}\vb*r'\bqty{\vb*J^c(\vb*r',t)\vdot\dd{\vb*S'}},
\label{momenttotapprox}
\end{equation}
as shown in the Appendix. Here \(\mathcal S_{reg}\) is the surface bounding the region. Substituting \eqref{momenttotapprox} into \eqref{momentsynapprox} gives
\begin{equation}
\vb*Q_{reg}^s(t)\approx\int\limits_{\mathcal S_{reg}}\vb*r'\bqty{\vb*J^c(\vb*r',t)\vdot\dd{\vb*S'}}-\vb*Q_{reg}^c(t).
\label{momentsynapproxsurfvol}
\end{equation}
Equivalently (see the Appendix), expression \eqref{momenttotapprox} can be written in the form
\begin{equation}
\vb*Q_{reg}(t)\approx\int\limits_{\mathcal V_{reg}}\vb*r'\grad'\vdot\vb*J^c(\vb*r',t)\dd[3]{r'}+\vb*Q_{reg}^c(t),
\label{momenttotapproxequiv}
\end{equation}
which can also be obtained independently by integrating Eq.~\eqref{currentdenscondimp} of the current density (see the Appendix). Importantly, \eqref{momenttotapprox} and \eqref{momenttotapproxequiv} are independent of the chosen coordinate origin if, and only if, the net impressed current crossing the region boundary is approximately zero (see the Appendix). Using \eqref{momentsynapprox} and \eqref{momenttotapproxequiv} we obtain
\begin{equation}
\vb*Q_{reg}^s(t)\approx\int\limits_{\mathcal V_{reg}}\vb*r'\grad'\vdot\vb*J^c(\vb*r',t)\dd[3]{r'}.
\label{momentsynapproxvol}
\end{equation}
We thus have two equivalent expressions, \eqref{momentsynapproxsurfvol} and \eqref{momentsynapproxvol}, of the synaptic current dipole moment. We observe that they are given in terms of the conduction current density. In the following we use its expression \eqref{condcurrentdens} to write \eqref{momentsynapproxsurfvol} and \eqref{momentsynapproxvol} in terms of the scalar potential.

\subsubsection{First expression in terms of the scalar potential}

Here we consider expression \eqref{momentsynapproxsurfvol}. By using \eqref{momentcond} and \eqref{condcurrentdens}, it takes the form
\begin{equation}
\begin{aligned}
\vb*Q_{reg}^s(t)&\approx-\int\limits_{\mathcal S_{reg}}\sigma(\vb*r',t)\vb*r'\bqty{\grad'\Phi(\vb*r',t)\vdot\dd{\vb*S'}}\\
&\quad+\int\limits_{\mathcal V_{reg}}\sigma(\vb*r',t)\grad'\Phi(\vb*r',t)\dd[3]{r'}.
\end{aligned}
\label{momentsynapproxsurfvolpot}
\end{equation}
The second term on the right-hand side here can be transformed as follows~\cite{geselowitz67}. The nonhomogeneity in the conductivity can be taken into account by visualizing the region as composed by \(m\) homogeneous parts with uniform conductivity \(\sigma_{reg}^{(i)}(t)\), where \(i=1, 2, \dots, m\). Representing the volume of the \(i\)th part and its bounding surface by \(\mathcal V_{reg}(i)\) and \(\mathcal S_{reg}(i)\), respectively, we have
\begin{equation}
\begin{aligned}
&\int\limits_{\mathcal V_{reg}}\sigma(\vb*r',t)\grad'\Phi(\vb*r',t)\dd[3]{r'}\\
&\enskip=\sum_{i=1}^m\sigma_{reg}^{(i)}(t)\int\limits_{\mathcal S_{reg}(i)}\Phi(\vb*r',t)\dd{\vb*S'_i},
\end{aligned}
\end{equation}
where the identity
\begin{equation}
\int\limits_{\mathcal V_{reg}(i)}\grad'\Phi(\vb*r',t)\dd[3]{r'}=\int\limits_{\mathcal S_{reg}(i)}\Phi(\vb*r',t)\dd{\vb*S'_i},
\end{equation}
is utilized. Here \(\dd{\vb*S'_i}\) is the outward normal vector of surface element for the \(i\)th part. Next, the surface \(\mathcal S_{reg}(i)\) can be divided into an internal component \(\mathcal S_{reg}^{int}(i)\) which lies inside the given region, and an external component \(\mathcal S_{reg}^{ext}(i)\) which makes part of the bounding surface \(\mathcal S_{reg}\) of the considered region. We then write
\begin{equation}
\begin{aligned}
\int\limits_{\mathcal S_{reg}(i)}\Phi(\vb*r',t)\dd{\vb*S'_i}&=\int\limits_{\mathcal S_{reg}^{int}(i)}\Phi(\vb*r',t)\dd{\vb*S'_i}\\
&\quad+\int\limits_{\mathcal S_{reg}^{ext}(i)}\Phi(\vb*r',t)\dd{\vb*S'_i}.
\end{aligned}
\end{equation}
Finally, representing by \(\mathcal S_{reg}^{int}(i,j)\) the common internal surface between two homogeneous parts \(i\) and \(j\), the current dipole moment \eqref{momentsynapproxsurfvolpot} is given by
\begin{equation}
\begin{aligned}
\vb*Q_{reg}^s(t)&\approx-\int\limits_{\mathcal S_{reg}}\sigma(\vb*r',t)\vb*r'\bqty{\grad'\Phi(\vb*r',t)\vdot\dd{\vb*S'}}\\
&\quad+\sum_{<ij>}\bqty{\sigma_{reg}^{(i)}(t)-\sigma_{reg}^{(j)}(t)}\\
&\qquad\qquad\times\int\limits_{\mathcal S_{reg}^{int}(i,j)}\Phi(\vb*r',t)\dd{\vb*S'_{i,j}}\\
&\quad+\int\limits_{\mathcal S_{reg}}\sigma(\vb*r',t)\Phi(\vb*r',t)\dd{\vb*S'}.
\end{aligned}
\label{momentsynapproxsurfspot}
\end{equation}
Here the sum is over all different pairs of adjacent homogeneous parts in the given region. The normal vector of surface element \(\dd{\vb*S'_{i,j}}\) is directed from the \(i\)th part to the \(j\)th part.

The above expression obtained for synaptic currents can be considered also valid for impressed currents. By integrating \eqref{impcurrentdens} and using \eqref{momentactapprox}, we see that the synaptic current dipole moment results to be approximately equal to the impressed current dipole moment. We note that an expression for this is given in Ref.~\citenum{geselowitz70}. However, such expression does not contain the first term on the right-hand side of \eqref{momentsynapproxsurfspot}. This term corresponds to the total current dipole moment \eqref{momenttotapprox}. In Ref.~\citenum{geselowitz70}, an external medium with zero conductivity is assumed implying that the normal component of the conduction current density vanishes on both sides close to the region boundary. In contrast, we assume an external medium with nonzero conductivity which corresponds to a non-isolated region. As such, the last expression obtained above can be seen as a generalization of the expression presented in Ref.~\citenum{geselowitz70}.

\subsubsection{Second expression in terms of the scalar potential}

Here we consider expression \eqref{momentsynapproxvol}. Using \eqref{condcurrentdens}, we have
\begin{equation}
\vb*Q_{reg}^s(t)\approx-\int\limits_{\mathcal V_{reg}}\vb*r'\grad'\vdot\bqty{\sigma(\vb*r',t)\grad'\Phi(\vb*r',t)}\dd[3]{r'}.
\end{equation}
Taking into account the nonhomogeneity in the conductivity as above, by considering the region as composed by \(m\) homogeneous parts with uniform conductivity \(\sigma_{reg}^{(i)}(t)\), where \(i=1, 2, \dots, m\), and representing the volume of the \(i\)th part by \(\mathcal V_{reg}(i)\), we get
\begin{equation}
\vb*Q_{reg}^s(t)\approx-\sum_{i=1}^m\sigma_{reg}^{(i)}(t)\int\limits_{\mathcal V_{reg}(i)}\vb*r'{\nabla'}^2\Phi(\vb*r',t)\dd[3]{r'}.
\end{equation}
Here, different from \eqref{momentsynapproxsurfspot}, the synaptic current dipole moment is in terms of the second spatial derivatives of the scalar potential inside each homogeneous part.

The last expression can also be written in terms of the free charge density \(\rho\). In the electro-quasistatic approximation we have \(\vb*E(\vb*r',t)\approx-\grad'\Phi(\vb*r',t)\). This leads to the Poisson's expression
\begin{equation}
{\nabla'}^2\Phi(\vb*r',t)\approx-\grad'\vdot\vb*E(\vb*r',t).
\end{equation}
Describing the \(i\)th part by a linear and isotropic medium with uniform permittivity \(\epsilon_{reg}^{(i)}(t)\), we have \(\vb*D(\vb*r',t)=\epsilon_{reg}^{(i)}(t)\vb*E(\vb*r',t)\). Considering the Maxwell's equation \eqref{divD}, we then obtain
\begin{equation}
\vb*Q_{reg}^s(t)\approx\sum_{i=1}^m\frac{\sigma_{reg}^{(i)}(t)}{\epsilon_{reg}^{(i)}(t)}\int\limits_{\mathcal V_{reg}(i)}\vb*r'\rho(\vb*r',t)\dd[3]{r'}.
\end{equation}
This can be written in a more compact manner by using
\begin{equation}
\tau_{reg}^{(i)}(t)=\frac{\epsilon_{reg}^{(i)}(t)}{\sigma_{reg}^{(i)}(t)},
\end{equation}
which is the Maxwell--Wagner time for the \(i\)th part, and
\begin{equation}
\vb*p_{reg}^{(i)}(t)=\int\limits_{\mathcal V_{reg}(i)}\vb*r'\rho(\vb*r',t)\dd[3]{r'},
\end{equation}
which is the electric dipole moment for the \(i\)th part. Then
\begin{equation}
\vb*Q_{reg}^s(t)\approx\sum_{i=1}^m\frac{\vb*p_{reg}^{(i)}(t)}{\tau_{reg}^{(i)}(t)}.
\end{equation}
Thus, through the corresponding Maxwell--Wagner time, the synaptic current dipole moment is related to the electric dipole moment of each homogeneous part.

\subsection{Long-distance magnetic field of action potential currents}

Here we consider the case of action potential currents, given by Eq.~\eqref{biotsavartBJtapproxmu0act}. Similarly to the synaptic case, we consider several relatively small regions in the medium, such that
\begin{equation}
\vb*B^a(\vb*r,t)=\sum_{reg}\vb*B_{reg}^a(\vb*r,t),
\label{Btactallregs}
\end{equation}
where
\begin{equation}
\vb*B_{reg}^a(\vb*r,t)=\frac{\mu_0^{\mathstrut}}{4\pi}\int\limits_{\mathcal V_{reg}}\vb*J^a(\vb*r',t)\cp\frac{\vb*r-\vb*r'}{\vqty{\vb*r-\vb*r'}^3}\dd[3]{r'},
\label{biotsavartBJtapproxmu0actreg}
\end{equation}
corresponds to a region of volume \(\mathcal V_{reg}\). As above, the dimensions of each region must be small compared to the average distance from the region to the field point \(\vb*r\). Taking any point \(\vb*r_{reg}^{\mathstrut}\) in the region, we must have
\begin{equation}
\vqty{\vb*r'-\vb*r_{reg}^{\mathstrut}}\ll\vqty{\vb*r-\vb*r_{reg}^{\mathstrut}},
\end{equation}
for every region point \(\vb*r'\).

Eq.~\eqref{biotsavartBJtapproxmu0actreg} can be written in the form
\begin{equation}
\vb*B_{reg}^a(\vb*r,t)=\curl\vb*A_{reg}^a(\vb*r,t),
\label{Bvectpotregact}
\end{equation}
with the vector potential
\begin{equation}
\vb*A_{reg}^a(\vb*r,t)=\frac{\mu_0^{\mathstrut}}{4\pi}\int\limits_{\mathcal V_{reg}}\frac{\vb*J^a(\vb*r',t)}{\vqty{\vb*r-\vb*r'}}\dd[3]{r'}.
\end{equation}
Employing the expansion \eqref{invdistexp} of the factor \(1/\vqty{\vb*r-\vb*r'}\) around \(\vb*r'=\vb*r_{reg}^{\mathstrut}\), and considering \eqref{momentactapprox}, we have
\begin{equation}
\begin{aligned}
\vb*A_{reg}^a(\vb*r,t)&\approx\frac{\mu_0^{\mathstrut}}{4\pi\vqty{\vb*r-\vb*r_{reg}^{\mathstrut}}^3}\\
&\quad\times\int\limits_{\mathcal V_{reg}}\vb*J^a(\vb*r',t)\bqty{\pqty{\vb*r-\vb*r_{reg}^{\mathstrut}}\vdot\vb*r'}\dd[3]{r'}.
\end{aligned}
\label{vectpotregexpact}
\end{equation}
By using \eqref{Bvectpotregact}, we then obtain
\begin{equation}
\begin{aligned}
\vb*B_{reg}^a(\vb*r,t)&\approx-\frac{3\mu_0^{\mathstrut}}{4\pi}\frac{\vu*R\pqty{\vb*r, \vb*r_{reg}^{\mathstrut}}}{\vqty{\vb*R\pqty{\vb*r, \vb*r_{reg}^{\mathstrut}}}^3}\\
&\quad\cp\int\limits_{\mathcal V_{reg}}\vb*J^a(\vb*r',t)\bqty{\vu*R\pqty{\vb*r, \vb*r_{reg}^{\mathstrut}}\vdot\vb*r'}\dd[3]{r'}\\
&\quad+\frac{\mu_0^{\mathstrut}}{2\pi}\frac{\vb*m_{reg}^a(t)}{\vqty{\vb*R\pqty{\vb*r, \vb*r_{reg}^{\mathstrut}}}^3},
\end{aligned}
\label{Bregact}
\end{equation}
where \(\vu*R\pqty{\vb*r, \vb*r_{reg}^{\mathstrut}}\) and \(\vb*R\pqty{\vb*r, \vb*r_{reg}^{\mathstrut}}\) are given by \eqref{relposunitvect} and \eqref{relposvect}, and
\begin{equation}
\vb*m_{reg}^a(t)=\frac12\int\limits_{\mathcal V_{reg}}\vb*r'\cp\vb*J^a(\vb*r',t)\dd[3]{r'},
\label{magnmomentact}
\end{equation}
is the magnetic dipole moment for the action potential currents. This, as well as the integral in \eqref{Bregact}, are independent of the chosen coordinate origin, which can be shown by using \eqref{momentactapprox}.

Expression \eqref{Bregact} shows a decrease with the cube of the distance for action potential currents. This is characteristic for the quadrupolar case~\cite{hamalainen}. Here we have a faster decrease with the distance than in the quadratic synaptic case given above by \eqref{Bregsyn}.

We note that expression \eqref{Bregact} can be written in a familiar form for a magnetic dipole, provided \(\divergence\vb*J^a\approx0\) in the given region, and \(\vb*J^a(\vb*r',t)=\vb0\) on its boundary~\cite{jackson,*wangsness}. In that case, the integral in \eqref{Bregact} can be expressed through the magnetic dipole moment \eqref{magnmomentact}, by
\begin{equation}
\begin{aligned}
&\int\limits_{\mathcal V_{reg}}\vb*J^a(\vb*r',t)\bqty{\vu*R\pqty{\vb*r, \vb*r_{reg}^{\mathstrut}}\vdot\vb*r'}\dd[3]{r'}\\
&\enskip\approx\vb*m_{reg}^a(t)\cp\vu*R\pqty{\vb*r, \vb*r_{reg}^{\mathstrut}},
\end{aligned}
\label{intBregact}
\end{equation}
and \eqref{Bregact} then takes the form
\begin{equation}
\begin{aligned}
\vb*B_{reg}^a(\vb*r,t)\approx\frac{\mu_0^{\mathstrut}}{4\pi}&\left\{3\vu*R\pqty{\vb*r, \vb*r_{reg}^{\mathstrut}}\right.\\
&\enskip\times\bqty{\vu*R\pqty{\vb*r, \vb*r_{reg}^{\mathstrut}}\vdot\vb*m_{reg}^a(t)}\\
&\left.\vphantom{\vu*R\pqty{\vb*r, \vb*r_{reg}^{\mathstrut}}}-\vb*m_{reg}^a(t)\right\}\frac1{\vqty{\vb*R\pqty{\vb*r, \vb*r_{reg}^{\mathstrut}}}^3}.
\end{aligned}
\label{Bregactmagndip}
\end{equation}
Additionally, using \eqref{intBregact}, the vector potential \eqref{vectpotregexpact} reads
\begin{equation}
\vb*A_{reg}^a(\vb*r,t)\approx\frac{\mu_0^{\mathstrut}}{4\pi}\vb*m_{reg}^a(t)\cp\frac{\vu*R\pqty{\vb*r, \vb*r_{reg}^{\mathstrut}}}{\vqty{\vb*R\pqty{\vb*r, \vb*r_{reg}^{\mathstrut}}}^2},
\end{equation}
which, through \eqref{Bvectpotregact}, can also be utilized to obtain \eqref{Bregactmagndip}.

\section{Discussion}

Here we have developed a method for calculating the long-distance magnetic field generated by neuronal populations. Since MEG measurements show magnetic signals that change over time, we first considered the time-dependent generalization of the Biot--Savart law given by Eq.~\eqref{Jefmko}. As the main contributions in neuromagnetism come from relatively low frequencies that are below a typical value \(f_{max}=\SI{100}{Hz}\)~\cite{hamalainen}, we then followed the approach of \citet{griffiths} to approximate Eq.~\eqref{Jefmko} in the quasistatic case. This method differs from previous ones~\cite{hamalainen,wagner,haus,gratiy} where the treatment does not rely on the general solution of Maxwell's equations, but on the quasistatic approximation of these equations themselves.

The time dependency of the current density and that of the magnetic field can be described via the different frequency components. We found that in the quasistatic case, the general nonstatic solution \eqref{Jefmko} can be approximated by expression \eqref{biotsavartBJtapprox} provided \(w_{max}^2t_{trav}^2(\vb*r,\bar{\vb*r}')\ll1\). This is expression \eqref{condgensol} and describes the quasistatic case through the current frequency component with the shortest oscillation period \(2\pi/w_{max}=1/f_{max}\). This time must be large enough compared to the largest traveling time \(t_{trav}(\vb*r,\bar{\vb*r}')\). The latter is the time needed for an electromagnetic signal to travel from the most distant source point \(\bar{\vb*r}'\) to the measurement point \(\vb*r\).

We note that a similar expression to \eqref{condgensol} is well known for the magneto-quasistatic approximation of Maxwell's equations~\cite{wagner,haus}, as well as their electro-quasistatic approximation~\cite{wagner,haus,gratiy}. In these cases, however, instead of the traveling distance \(\vqty{\vb*r-\bar{\vb*r}'}\), one typical or characteristic length scale of the medium is employed. The system dimensions are assumed to be all within a factor of two or so of each other~\cite{haus}. The typical length of gray matter is estimated by the cortical thickness~\cite{wagner,gratiy}. On the other hand, Eq.~\eqref{Jefmko} leads to expression \eqref{condgensol} and therefore, to the traveling distance \(\vqty{\vb*r-\bar{\vb*r}'}\) as the appropriate length to define the quasistatic dynamics.

A clearly different expression from \eqref{condgensol} can be found in Ref.~\citenum{hamalainen} where the quasistatic approximation of the Ampère--Maxwell equation was examined. It is given by \(w_{max}\tau_{MW}^{\mathstrut}\ll1\), where \(\tau_{MW}^{\mathstrut}=\epsilon/\sigma\) is the Maxwell--Wagner reaction time. That is expression \eqref{condAmpMaxw} which we discussed in greater detail earlier. Here \(\epsilon\) and \(\sigma\) are the permittivity and the electrical conductivity of the medium, respectively. We note that a permittivity \(\epsilon\approx\num{e5}\epsilon_0^{\mathstrut}\), where \(\epsilon_0^{\mathstrut}\) is the permittivity of free space, was considered in Ref.~\citenum{hamalainen}. This gives \(w_{max}\tau_{MW}^{\mathstrut}\approx\num{2e-3}\ll1\). Here, however, based on Refs.~\citenum{wagner} and \citenum{gabriel}, the permittivity for gray matter at frequencies below \SI{100}{Hz} has been underestimated by two orders of magnitude. Therefore, the condition \(w_{max}\tau_{MW}^{\mathstrut}\ll1\) is actually fulfilled in a minimal manner.

The quasistatic approximation \eqref{biotsavartBJtapprox} of the Biot--Savart law was justified with \eqref{condgensol} being well satisfied in the form \((2\pi f_{max})^2\vqty{\vb*r-\bar{\vb*r}'}^2\epsilon\mu\approx\num{4e-7}\ll1\). Here we estimated \(\vqty{\vb*r-\bar{\vb*r}'}\approx\SI{.1}{m}\), mainly given by the average size of the human brain. Additionally, we took the permittivity \(\epsilon\approx\num{e7}\epsilon_0^{\mathstrut}\) for gray matter at frequencies below \SI{100}{Hz}~\cite{wagner,gabriel}. Finally, typical for biological tissues~\cite{wagner}, the permeability \(\mu\) was approximated by that of free space, \(\mu_0^{\mathstrut}\).

We then focused on the magnetic field of the various currents. \alain{Following previous work~\cite{geselowitz67,ilmoniemi},} the current density was split into conduction and impressed current densities. The conduction current is caused by the electric field and its density is specified in the electro-quasistatic approximation by \(\vb*J^c(\vb*r',t)\approx-\sigma(\vb*r',t)\grad'\Phi(\vb*r',t)\), where \(\sigma(\vb*r',t)\) is the nonhomogeneous conductivity of the medium and \(\Phi(\vb*r',t)\) is the scalar potential. We assumed the impressed current density mostly given by a synaptic current density \(\vb*J^s(\vb*r',t)\) and an action potential current density \(\vb*J^a(\vb*r',t)\). The magnetic field case of conduction currents has been worked out exactly~\cite{geselowitz70}. The long-distance magnetic field was studied in the synaptic and action potential cases.

Each population of neurons was described as a relatively small region such that its dimensions are small compared to the average distance from the region to the field point \(\vb*r\). Taking any point \(\vb*r_{reg}^{\mathstrut}\) in the region, we must have \(\vqty{\vb*r'-\vb*r_{reg}^{\mathstrut}}\ll\vqty{\vb*r-\vb*r_{reg}^{\mathstrut}}\), for every region point \(\vb*r'\). Considering the distance to the measurement point in the order of centimeters, each region volume \(\mathcal V_{reg}\) can be thought in the order of \SI{1}{mm^3}. Using known values of the cortical surface, the average cortical thickness, and the total number of neurons in the human cerebral cortex~\cite{essen}, we can find a quantity of approximately \num{4.7e5} regions of \SI{1}{mm^3} in the human cerebral cortex, each one with an average estimate of about \num{3.4e4} neurons. The contributions from the different regions must be added together to obtain the magnetic field at a certain point. This is indicated through Eqs.~\eqref{Btsynallregs} and \eqref{Btactallregs} for synaptic currents and action potential currents, respectively.

The magnetic induction produced by synaptic currents at a relatively distant point \(\vb*r\), for a region located at \(\vb*r_{reg}^{\mathstrut}\), can be approximated by expression \eqref{Bregsyn}. Here \(\vb*Q_{reg}^s(t)\) is the (time-dependent) synaptic current dipole moment defined by Eq.~\eqref{momentsyn}. This was related to the conduction current density leading to two equivalent expressions of \(\vb*Q_{reg}^s(t)\) in terms of the scalar potential.

\alain{In conclusion, we have derived here the formalism to relate the magnetic field at large distances (centimeters) from populations of neurons and estimated the contributions from various current sources in neurons.   The next step in this approach is to apply this population-level description to mean-field models of neuronal populations, in order to link these models with the genesis of the MEG signal.}

\appendix*

\section{Total current dipole moment}

Here we derive expressions \eqref{momenttotapprox} and \eqref{momenttotapproxequiv} for the total current dipole moment. It is also checked how each of these expressions can yield the same result regardless of the choice of the coordinate origin. The following are two key assumptions. Firstly, we consider quasistatic conditions such that the continuity equation gives
\begin{equation}
\grad'\vdot\vb*J(\vb*r',t)=-\pdv{\rho(\vb*r',t)}{t}\approx0.
\label{conteqquasistat}
\end{equation}
Secondly, we assume a small contribution of the impressed currents at the surface \(\mathcal S_{reg}\) bounding the given region, compared to that of the conduction currents. We take
\begin{equation}
\vqty{\,\int\limits_{\mathcal S_{reg}}x'\vb*J^i(\vb*r',t)\vdot\dd{\vb*S'}}\ll\vqty{\,\int\limits_{\mathcal S_{reg}}x'\vb*J^c(\vb*r',t)\vdot\dd{\vb*S'}},
\label{rprimpcurrentdenssurfintcompx}
\end{equation}
and this similarly applies to the remaining components of \(\int\limits_{\mathcal S_{reg}}\vb*r'\bqty{\vb*J^i(\vb*r',t)\vdot\dd{\vb*S'}}\).

\subsection{Expression \eqref{momenttotapprox}}

We consider the \(x\) component of the total current dipole moment \eqref{momenttot}, given by
\begin{equation}
Q_{reg}^x(t)=\int\limits_{\mathcal V_{reg}}J_x(\vb*r',t)\dd[3]{r'},
\end{equation}
and write \(J_x(\vb*r',t)\) in the form
\begin{equation}
\begin{aligned}
J_x(\vb*r',t)=\vb*J(\vb*r',t)\vdot\grad'x'&=\grad'\vdot[x'\vb*J(\vb*r',t)]\\
&\quad-x'\grad'\vdot\vb*J(\vb*r',t).
\end{aligned}
\label{currentdenscompx}
\end{equation}
By using \eqref{conteqquasistat}, this simplifies to
\begin{equation}
J_x(\vb*r',t)\approx\grad'\vdot[x'\vb*J(\vb*r',t)].
\end{equation}
Then
\begin{equation}
\begin{aligned}
Q_{reg}^x(t)&\approx\int\limits_{\mathcal V_{reg}}\grad'\vdot[x'\vb*J(\vb*r',t)]\dd[3]{r'}\\
&=\int\limits_{\mathcal S_{reg}}x'\vb*J(\vb*r',t)\vdot\dd{\vb*S'},
\end{aligned}
\end{equation}
where \(\mathcal S_{reg}\) is the surface bounding the region, according to the divergence theorem. Furthermore, from Eq.~\eqref{currentdenscondimp}, \(\vb*J(\vb*r',t)=\vb*J^c(\vb*r',t)+\vb*J^i(\vb*r',t)\). Taking then into account \eqref{rprimpcurrentdenssurfintcompx}, we write
\begin{equation}
Q_{reg}^x(t)\approx\int\limits_{\mathcal S_{reg}}x'\vb*J^c(\vb*r',t)\vdot\dd{\vb*S'}.
\label{momenttotapproxappcompx}
\end{equation}

Considering the \(y\) and \(z\) components of the total current dipole moment, we similarly obtain
\begin{equation}
Q_{reg}^y(t)\approx\int\limits_{\mathcal S_{reg}}y'\vb*J^c(\vb*r',t)\vdot\dd{\vb*S'},
\end{equation}
and
\begin{equation}
Q_{reg}^z(t)\approx\int\limits_{\mathcal S_{reg}}z'\vb*J^c(\vb*r',t)\vdot\dd{\vb*S'}.
\end{equation}
Therefore, we get
\begin{equation}
\vb*Q_{reg}(t)\approx\int\limits_{\mathcal S_{reg}}\vb*r'\bqty{\vb*J^c(\vb*r',t)\vdot\dd{\vb*S'}},
\label{momenttotapproxapp}
\end{equation}
that is, expression \eqref{momenttotapprox}.

\subsection{Expression \eqref{momenttotapproxequiv}}

\subsubsection{Derivation from expression \eqref{momenttotapprox}}

Considering the \(x\) component in \eqref{momenttotapprox} (or \eqref{momenttotapproxapp}), given by \eqref{momenttotapproxappcompx}, we have
\begin{equation}
Q_{reg}^x(t)\approx\int\limits_{\mathcal V_{reg}}\grad'\vdot[x'\vb*J^c(\vb*r',t)]\dd[3]{r'},
\end{equation}
using the divergence theorem. Similarly to \eqref{currentdenscompx}, we get
\begin{equation}
\grad'\vdot[x'\vb*J^c(\vb*r',t)]=x'\grad'\vdot\vb*J^c(\vb*r',t)+J_x^c(\vb*r',t).
\end{equation}
Then
\begin{equation}
\begin{aligned}
Q_{reg}^x(t)&\approx\int\limits_{\mathcal V_{reg}}x'\grad'\vdot\vb*J^c(\vb*r',t)\dd[3]{r'}\\
&\quad+\int\limits_{\mathcal V_{reg}}J_x^c(\vb*r',t)\dd[3]{r'}.
\end{aligned}
\label{momenttotapproxequivappcompx}
\end{equation}
The second term on the right-hand side here is the \(x\) component of the conduction current dipole moment, \(\vb*Q_{reg}^c(t)\), defined by \eqref{momentcond}. Expressions of the same form as that of \eqref{momenttotapproxequivappcompx} apply to the \(y\) and \(z\) components. Therefore
\begin{equation}
\vb*Q_{reg}(t)\approx\int\limits_{\mathcal V_{reg}}\vb*r'\grad'\vdot\vb*J^c(\vb*r',t)\dd[3]{r'}+\vb*Q_{reg}^c(t),
\end{equation}
that is, we obtain expression \eqref{momenttotapproxequiv}.

\subsubsection{Independent derivation}

Expression \eqref{momenttotapproxequiv} can also be derived starting from Eq.~\eqref{currentdenscondimp} of the current density. Integrating it, we have
\begin{equation}
\vb*Q_{reg}(t)=\vb*Q_{reg}^c(t)+\vb*Q_{reg}^i(t),
\label{currentdenscondimpint}
\end{equation}
where
\begin{equation}
\vb*Q_{reg}^i(t)=\int\limits_{\mathcal V_{reg}}\vb*J^i(\vb*r',t)\dd[3]{r'},
\end{equation}
is the impressed current dipole moment, with \(x\) component
\begin{equation}
\begin{aligned}
\int\limits_{\mathcal V_{reg}}J_x^i(\vb*r',t)\dd[3]{r'}&=\int\limits_{\mathcal V_{reg}}\vb*J^i(\vb*r',t)\vdot\grad'x'\dd[3]{r'}\\
&=\int\limits_{\mathcal V_{reg}}\grad'\vdot[x'\vb*J^i(\vb*r',t)]\dd[3]{r'}\\
&\quad-\int\limits_{\mathcal V_{reg}}x'\grad'\vdot\vb*J^i(\vb*r',t)\dd[3]{r'}.
\end{aligned}
\end{equation}
Then
\begin{equation}
\begin{aligned}
\int\limits_{\mathcal V_{reg}}J_x^i(\vb*r',t)\dd[3]{r'}&=\int\limits_{\mathcal S_{reg}}x'\vb*J^i(\vb*r',t)\vdot\dd{\vb*S'}\\
&\quad-\int\limits_{\mathcal V_{reg}}x'\grad'\vdot\vb*J^i(\vb*r',t)\dd[3]{r'}.
\end{aligned}
\label{momentimpappcompx}
\end{equation}
Considering again \eqref{currentdenscondimp}, we have
\begin{equation}
\grad'\vdot\vb*J(\vb*r',t)=\grad'\vdot\vb*J^c(\vb*r',t)+\grad'\vdot\vb*J^i(\vb*r',t),
\end{equation}
which through \eqref{conteqquasistat}, implies
\begin{equation}
\grad'\vdot\vb*J^i(\vb*r',t)\approx-\grad'\vdot\vb*J^c(\vb*r',t).
\label{conteqquasistatimpcond}
\end{equation}
Thus
\begin{equation}
\begin{aligned}
\int\limits_{\mathcal V_{reg}}J_x^i(\vb*r',t)\dd[3]{r'}&\approx\int\limits_{\mathcal S_{reg}}x'\vb*J^i(\vb*r',t)\vdot\dd{\vb*S'}\\
&\quad+\int\limits_{\mathcal V_{reg}}x'\grad'\vdot\vb*J^c(\vb*r',t)\dd[3]{r'}.
\end{aligned}
\end{equation}
Here we can substitute
\begin{equation}
\begin{aligned}
\int\limits_{\mathcal V_{reg}}x'\grad'\vdot\vb*J^c(\vb*r',t)\dd[3]{r'}&=\int\limits_{\mathcal S_{reg}}x'\vb*J^c(\vb*r',t)\vdot\dd{\vb*S'}\\
&\quad-\int\limits_{\mathcal V_{reg}}J_x^c(\vb*r',t)\dd[3]{r'},
\end{aligned}
\end{equation}
which is similar to \eqref{momentimpappcompx}. Using then \eqref{rprimpcurrentdenssurfintcompx}, we get
\begin{equation}
\int\limits_{\mathcal V_{reg}}J_x^i(\vb*r',t)\dd[3]{r'}\approx\int\limits_{\mathcal V_{reg}}x'\grad'\vdot\vb*J^c(\vb*r',t)\dd[3]{r'}.
\end{equation}
Expressions of the same form hold for the \(y\) and \(z\) components of the impressed current dipole moment. Therefore
\begin{equation}
\vb*Q_{reg}^i(t)\approx\int\limits_{\mathcal V_{reg}}\vb*r'\grad'\vdot\vb*J^c(\vb*r',t)\dd[3]{r'}.
\end{equation}
Eq.~\eqref{currentdenscondimpint} then gives
\begin{equation}
\vb*Q_{reg}(t)\approx\int\limits_{\mathcal V_{reg}}\vb*r'\grad'\vdot\vb*J^c(\vb*r',t)\dd[3]{r'}+\vb*Q_{reg}^c(t),
\label{momenttotapproxequivapp}
\end{equation}
thus reproducing expression \eqref{momenttotapproxequiv}.

\subsection{Independence on the coordinate origin}

The total current dipole moment
\begin{equation}
\vb*Q_{reg}(t)=\int\limits_{\mathcal V_{reg}}\vb*J(\vb*r',t)\dd[3]{r'},
\end{equation}
is independent of the chosen origin of the coordinate system. The current density value at a given point clearly does not depend on the position of the coordinate origin.

The independence of \(\vb*Q_{reg}(t)\) on the coordinate origin can be explicitly shown as follows. Let a new origin be set at position \(\vb*{\mathcal R}\) relative to the old origin. A given point located at \(\vb*r'\) in relation to the old origin is now located at \(\vb*r''\) in relation to the new origin, with \(\vb*r''=\vb*r'-\vb*{\mathcal R}\). Correspondingly, the current density is described by \(\vb*{\mathcal J}(\vb*r'',t)\), with \(\vb*{\mathcal J}(\vb*r'',t)=\vb*J(\vb*{\mathcal R}+\vb*r'',t)=\vb*J(\vb*r',t)\). The total current dipole moment when using the new origin is
\begin{equation}
\begin{aligned}
\vb*{\mathcal Q}_{reg}(t)&=\int\limits_{\mathcal V_{reg}}\vb*{\mathcal J}(\vb*r'',t)\dd[3]{r''}\\
&=\int\limits_{\mathcal V_{reg}}\vb*J(\vb*{\mathcal R}+\vb*r'',t)\dd[3]{r''},
\end{aligned}
\end{equation}
so that
\begin{equation}
\vb*{\mathcal Q}_{reg}(t)=\int\limits_{\mathcal V_{reg}}\vb*J(\vb*r',t)\dd[3]{r'}=\vb*Q_{reg}(t).
\end{equation}

The mentioned independence on the coordinate origin must be verified for expressions \eqref{momenttotapprox} and \eqref{momenttotapproxequiv} of the total current dipole moment. Below, when using a new origin as above, the conduction current density is described by
\begin{equation}
\vb*{\mathcal J}^c(\vb*r'',t)=\vb*J^c(\vb*{\mathcal R}+\vb*r'',t)=\vb*J^c(\vb*r',t).
\end{equation}

\subsubsection{First case}

Regarding expression \eqref{momenttotapprox} (or \eqref{momenttotapproxapp}), we have
\begin{equation}
\begin{aligned}
\vb*{\mathcal Q}_{reg}(t)&\approx\int\limits_{\mathcal S_{reg}}\vb*r''\bqty{\vb*{\mathcal J}^c(\vb*r'',t)\vdot\dd{\vb*S''}}\\
&=\int\limits_{\mathcal S_{reg}}\pqty{\vb*r'-\vb*{\mathcal R}}\bqty{\vb*J^c(\vb*r',t)\vdot\dd{\vb*S'}},
\end{aligned}
\end{equation}
so that
\begin{equation}
\begin{aligned}
\vb*{\mathcal Q}_{reg}(t)&\approx\int\limits_{\mathcal S_{reg}}\vb*r'\bqty{\vb*J^c(\vb*r',t)\vdot\dd{\vb*S'}}\\
&\quad-\vb*{\mathcal R}\int\limits_{\mathcal S_{reg}}\vb*J^c(\vb*r',t)\vdot\dd{\vb*S'}.
\end{aligned}
\end{equation}
Concerning the second term on the right-hand side here, by integrating \eqref{conteqquasistatimpcond} we get
\begin{equation}
\int\limits_{\mathcal S_{reg}}\vb*J^c(\vb*r',t)\vdot\dd{\vb*S'}\approx-\int\limits_{\mathcal S_{reg}}\vb*J^i(\vb*r',t)\vdot\dd{\vb*S'}.
\label{conteqquasistatimpcondint}
\end{equation}
Hence, for expression \eqref{momenttotapprox} to be independent of the coordinate origin, it is enough, and necessary, that the net impressed current crossing the region boundary is approximately zero. Namely, if, and only if, we have
\begin{equation}
\int\limits_{\mathcal S_{reg}}\vb*J^i(\vb*r',t)\vdot\dd{\vb*S'}\approx0,
\label{impcurrentdenssurfint}
\end{equation}
then
\begin{equation}
\int\limits_{\mathcal S_{reg}}\vb*J^c(\vb*r',t)\vdot\dd{\vb*S'}\approx0,
\end{equation}
and consequently
\begin{equation}
\vb*{\mathcal Q}_{reg}(t)\approx\int\limits_{\mathcal S_{reg}}\vb*r'\bqty{\vb*J^c(\vb*r',t)\vdot\dd{\vb*S'}},
\end{equation}
which gives the same result as expression \eqref{momenttotapprox}.

\subsubsection{Second case}

As for expression \eqref{momenttotapproxequiv} (or \eqref{momenttotapproxequivapp}), when we employ a new origin it reads
\begin{equation}
\begin{aligned}
\vb*{\mathcal Q}_{reg}(t)&\approx\int\limits_{\mathcal V_{reg}}\vb*r''\grad''\vdot\vb*{\mathcal J}^c(\vb*r'',t)\dd[3]{r''}\\
&\quad+\int\limits_{\mathcal V_{reg}}\vb*{\mathcal J}^c(\vb*r'',t)\dd[3]{r''},
\end{aligned}
\end{equation}
which, using the divergence theorem and \eqref{conteqquasistatimpcondint}, leads to
\begin{equation}
\begin{aligned}
\vb*{\mathcal Q}_{reg}(t)&\approx\int\limits_{\mathcal V_{reg}}\vb*r'\grad'\vdot\vb*J^c(\vb*r',t)\dd[3]{r'}+\vb*Q_{reg}^c(t)\\
&\quad+\vb*{\mathcal R}\int\limits_{\mathcal S_{reg}}\vb*J^i(\vb*r',t)\vdot\dd{\vb*S'}.
\end{aligned}
\end{equation}
Thus, similar to the case of expression \eqref{momenttotapprox}, if, and only if, \eqref{impcurrentdenssurfint} is met, we have
\begin{equation}
\vb*{\mathcal Q}_{reg}(t)\approx\int\limits_{\mathcal V_{reg}}\vb*r'\grad'\vdot\vb*J^c(\vb*r',t)\dd[3]{r'}+\vb*Q_{reg}^c(t),
\end{equation}
which produces the same result as expression \eqref{momenttotapproxequiv}.

\begin{acknowledgments}

Research funded by the CNRS, the European Community (H2020-785907), the ANR (PARADOX) and the ICODE excellence network.

\end{acknowledgments}

\bibliography{references}

\begin{thebibliography}{26}
\bibitem [{gra()}]{gratiyadd}%
  \BibitemOpen
  \href@noop {} {}\bibinfo {note} {{h}ere the approach is based on
  Ref.~\citenum{haus}}\BibitemShut {NoStop}%
\end{thebibliography}

\end{document}